# Interpretable Classification from Skin Cancer Histology Slides Using Deep Learning: A Retrospective Multicenter Study

Peizhen Xie[1]     Ke Zuo[1]     Yu Zhang[2*]     Fangfang Li[2]     Mingzhu Yin[3]     Kai Lu[1]

[1]National University of Defense Technology
[2]Xiangya Hospital, Central South University
[3]Yale School of Medicine

## Abstract

*For diagnosing melanoma, hematoxylin and eosin (H&E) stained tissue slides remains the gold standard. These images contain quantitative information in different magnifications. In the present study, we investigated whether deep convolutional neural networks can extract structural features of complex tissues directly from these massive size images in a patched way. In order to face the challenge arise from morphological diversity in histopathological slides, we built a multicenter database of 2241 digital whole-slide images from 1321 patients from 2008 to 2018. We trained both ResNet50 and Vgg19 using over 9.95 million patches by transferring learning, and test performance with two kinds of critical classifications: malignant melanomas versus benign nevi in separate and mixed magnification; and distinguish among nevi in maximum magnification. The CNNs achieves superior performance across both tasks, demonstrating an AI capable of classifying skin cancer in the analysis from histopathological images. For making the classifications reasonable, the visualization of CNN representations is furthermore used to identify cells between melanoma and nevi. Regions of interest (ROI) are also located which are significantly helpful, giving pathologists more support of correctly diagnosis.*

## 1. Introduction

Melanoma is a cancer that begins in the melanocytes, typically occur in the skin, but may rarely occur in the mouth, intestines, or eye [1-3]. Melanoma has an early metastasis associated with high degree of malignancy and mortality [2].

Although the prognosis of patients with advanced melanoma varies in different countries, early diagnosis can significantly reduce mortality [4, 5]. Currently routine examination of hematoxylin and eosin (H&E) stained tissue sections remains the gold standard in diagnosing melanoma. However, this qualitative process is not only time-consuming, but also mostly dependent on the experience level of clinician. A critical shortage of pathologists in China is hindering the ability of medical laboratories in the region to properly diagnose and classify diseases. It is urgent to continuously innovate in the auxiliary diagnosis technology such as automated classification and interpretable diagnosis through medical imaging to alleviate the problem of diagnosis bottleneck.

Convolutional neural networks (CNNs) [6-9, 22] have achieved superior performance over standard in various computer vision tasks. However, the end-to-end learning strategy makes the entire CNN a black box, which is difficult for people to understand the logic inside a CNN. Especially in the medical field, this prominent gap must be closed from research to clinical practice. In recent years, a growing number of researchers have realized that significant value of interpretability is in both theory and practice; various attempts have been made with interpretable knowledge representations [10-13]. Visualization of filters in a CNN is the most direct way of exploring visual patterns hidden inside a neural unit and widely used in interpretability of image-based diagnose. Daniel *et al.* [14] developed an artificial intelligence system using transfer learning. It effectively classified images for macular degeneration and diabetic retinopathy and also accurately distinguished bacterial and viral pneumonia on chest X-rays. Otherwise, it performed a visualization test to identify the ROI contributing most to the neural network's assignment of the predicted diagnosis. Arunima *et al.* [15] proposed a workflow which intelligently samples the training data by automatically selecting only image areas that display visible disease-relevant tissue state and isolates regions most pertinent to the trained CNN prediction and translates them to observable and qualitative features such as color, intensity, cell and tissue morphology and texture. This enhanced CNN based workflow both increased patient attribute predictive accuracy and experimentally proved that a data-driven CNN histology model predicting breast invasive carcinoma stages is highly sensitive to those found features. Ian *et al.* [16] present a novel spatial algorithm for assessing glaucoma in images of the optic nerve interpretable automated spatial analysis of the whole cup to disc profile. Seong *et al.*[17] designed visually interpretable diagnosis network which can focus important areas encoded in a guide map for diagnostic decisions. Samuel *et al.* [18] present a deep neural network model which enables classification of out-of-focus microscope images with both higher accuracy and greater precision via interpretable patch-level focus.

---

\* Corresponding author: Dr. Yu Zhang. Email: zhyu@csu.edu.cn.
This work was supported by "The Project of Science and Technology Leading Talent"-1404060117001.

Although artificial intelligence technology has achieved remarkable results in the field of medical imaging, it is still in its infancy for skin pathology analysis. The main reason is that the long time required for pathology image collection, in addition with doubts about the ability of deep learning on more complicated histopathology. In order to answer these questions, this paper specifically designed simulation experiments, which are close to the clinical diagnosis process, to evaluate the recognition ability of convolutional networks, and tried to make the their decisions interpretable. The main contributions of this paper can be summarized as: 1) To best of our knowledge, we for the first time used the visualization of CNN representations to distinguish between melanoma cells and nevus cells, and then to accurately locate ROI, which are helpful for assisting doctors in diagnosis. 2) We demonstrate classification of skin diseases using both ResNet50 [8] and Vgg19 [19] by transfer learning mechanisms, using only pixels and disease labels as inputs. Extensive experiments simulating the clinical diagnosis process are conducted to evaluate the effectiveness. 3) We built a multicenter database of 2241 whole-slide images from 1321 patients from 2008 to 2018 with annotations. These different magnification slides are in turn patched into over 9.95 million training images. This data-driven approach can overcome the challenge arise from morphological diversity in histopathological images.

## 2. Dataset

Our dataset is composed of 2241 H&E stained whole-slide histopathology images pathologist-labeled and organized of 4 diseases which are melanoma (MM), intradermal nevus (IN), compound nevus (CN) and junctional nevus (JN) respectively. The images come from the NCI's Genomic Data Commons which is clinician-curated, open-access online repository, as well as from clinical data from Central South University Xiangya Hospital, and Tissue Micro-Array center Yale School of

Medicine. These slides contain 8350 tissues, and each labeled tissue is made into over 9.95 million patches with size 256×256 in four different magnifications, which are 4X(5 microns/pixel), 10X (1 microns/pixel), 20X(0.5 microns/pixel) and 40X (0.275 microns/pixel). Figure 1 shows patch examples of each disease in different magnifications. In this paper we represent the result of two CNNs that matches the performance of pathologists at four kinds of key diagnostic tasks: melanoma classification from nevi in four different magnifications, melanoma classification from each kind of nevus in 40X, melanoma classification from nevi in mixed magnifications, and nevi classifications.

## 3. Methodology

A Convolutional Neural Network (CNN) are a special kind of multi-layer neural networks, designed to recognize visual patterns directly from pixel images with minimal preprocessing. Vgg19 consists of 19 convolutional layers and is very appealing because of its very uniform architecture. It is currently the most preferred choice in the community for extracting features from images. The so-called Residual Neural Network (ResNet) introduced a novel architecture with "skip connections" and features heavy batch normalization. Thanks to this technique they were able to train a NN with deeper layers while still having lower complexity than VggNet. In our paper, we chose Vgg19 and ResNet50 to test these four kinds of critical classification tasks and compared their performance.

Figure 3 shows our working system. After slides collection, the ground truth was obtained via manual delineation of the tissue region by 10 dermatologists and 20 pathologists. From the data with ground truth, we split 70% data as training set, 15% as the validation set, and 15% as testing set. To assess the performance on WSIs, the per-patch results were aggregated on one slide.

With image classification tasks, especially in medical field, we are often faced a challenge to explain our reasoning by dissecting the image, and pointing out prototypical aspects of one class or another. The visualization of CNN representations is the most direct way to explore network representations. The network visualization also provides a technical foundation for many approaches to diagnosing CNN representations. In our paper, explanations of individual network decisions have been explored by generating informative heatmaps such as CAM [20] and grad-CAM [21].

## 4. Experiments

In this section, we will first describe the evaluation metric used in our experiments and then demonstrate the performances of different methods.

## 4.1. Evaluation Metric

The experiment employs several metrics for performance evaluation, which includes precision (P), recall (R), F1_score (F1), sensitivity (SE) and specificity (SP). Let $N_{tp}$, $N_{tn}$, $N_{fp}$ and $N_{fn}$ represent the number of true positive, true negative, false positive and false negative, respectively, the criteria can be defined as:

$$P = \frac{N_{tp}}{N_{tp} + N_{fp}} \tag{1}$$

$$R = \frac{N_{tp}}{N_{tp} + N_{fn}} \tag{2}$$

$$F1 = \frac{\left(recall^{-1} + precision^{-1}\right)^{-1}}{2} = 2 \cdot \frac{precision \cdot reall}{precision + reall} \tag{3}$$

$$SE = \frac{N_{tp}}{N_{tp} + N_{tp}} \tag{4}$$

$$SP = \frac{N_{tn}}{N_{tn} + N_{fp}} \tag{5}$$

These evaluation metrics, i.e. SE and SP, are employed to assess the performance of pathological lesion classification. The primary metric ranking the results for these two tasks is the area under the ROC curve, i.e. AUC, which is generated by evaluating the true positive rate (TPR), i.e. SE, against the false positive rate (FPR), defined in (6), at various threshold settings.

$$FPR = \frac{N_{fp}}{N_{tn} + N_{fp}} = 1\text{-}SP \tag{6}$$

## 4.2. Implementation details

### 4.2.1. Melanoma from Nevi in four magnifications

In first evaluation, we need to identify the ability for classification of melanoma from three nevi, which are intradermal nevus (IN), junctional nevus (JN) and compound nevus (CN). In order to compare performance in different magnifications, we separately test F1/Sensitivity/Specificity/AUC on both ResNet50 and Vgg19 models in a strategy of transferring learning. Data sizes consist of 49533 patches from 4X, 104816 patches from 10X, 101070 patches from 20X and 101034 patches from 40X. Table 1 presents each metric in four magnifications. We found that best values of each metric are shown in 40X of both models. In each magnification, the performance values of ResNet50 are all better than Vgg19. Relatively, the value of F1 can be reached at 0.93, sensitivity 0.92, specificity 0.97 and AUC 0.99.

Table 1 Performance comparison of two CNNs models in four magnifications. The higher value is better. The results highlighted with black bold show the best performance in our dataset.

| Magnification | Model | F1 | Sensitivity | Specificity | AUC |
|---|---|---|---|---|---|
| 4X | ResNet50 | **0.87** | **0.92** | **0.93** | **0.98** |
|  | Vgg19 | 0.80 | 0.83 | 0.90 | 0.95 |
| 10X | ResNet50 | **0.91** | **0.90** | **0.97** | **0.98** |
|  | Vgg19 | 0.85 | 0.85 | 0.94 | 0.96 |
| 20X | ResNet50 | **0.90** | **0.88** | **0.97** | **0.98** |
|  | Vgg19 | 0.85 | 0.86 | 0.91 | 0.96 |
| 40X | ResNet50 | **0.93** | **0.92** | **0.97** | **0.99** |
|  | Vgg19 | 0.87 | 0.86 | 0.94 | 0.96 |

### 4.2.2 Melanoma from Nevi in Mixed magnifications

In a clinical scenario, pathologists often use images in a manner of mixed magnification for cancer diagnosis. Therefore, we designed this evaluation for AI to simulate the process. In this test, patches of each magnification are randomly selected data source to build this dataset. Data sizes consist 40928 patches of IN, 43430 patches of JN, 48284 patches of CN and 69020 patches from MM. As shown in table 2, the same conclusion can be reached that the values of ResNet50 in each metric are better than the values of Vgg19. Relatively, the value of F1 can reached at 0.89, sensitivity 0.92, specificity 0.94 and AUC 0.98.

Table 2 Performance comparison of two CNNs models in mixed magnifications. The higher value is better. The results highlighted with black bold show the best performance in our dataset.

| | Model | F1 | Sensitivity | Specificity | AUC |
|---|---|---|---|---|---|
| melanoma | ResNet50 | **0.89** | **0.92** | **0.94** | **0.98** |
|  | Vgg19 | 0.80 | 0.84 | 0.89 | **0.95** |

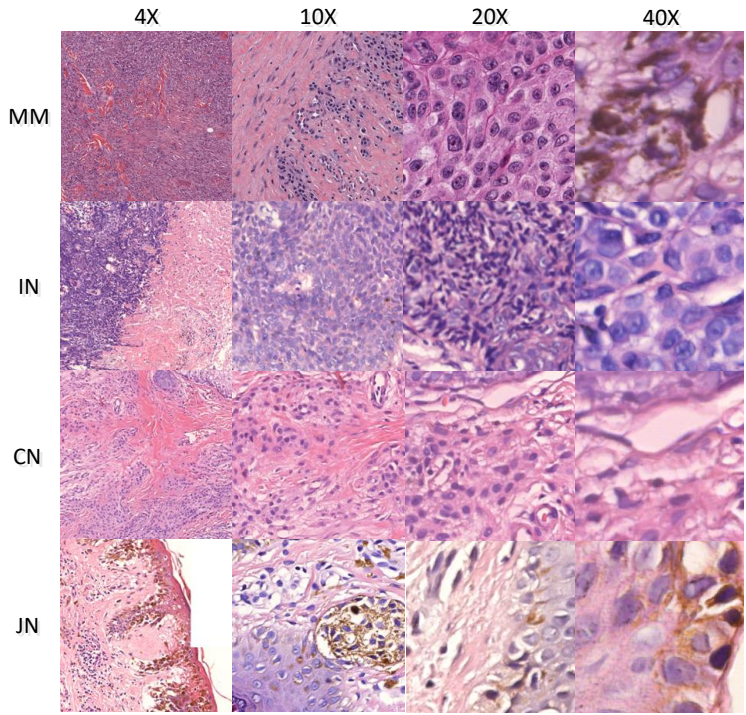

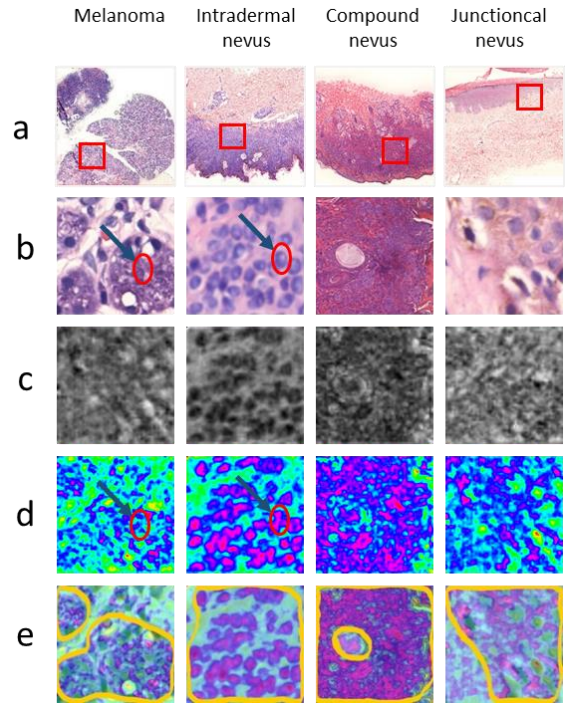

Figure 1 Some sample patches in different magnifications from dataset. Different magnifications are also illustrated in each disease. The abbreviation MM stands for melanoma, IN stands for intradermal nevus, JN stands for junctional nevus and CN stands for compound nevus.

Figure 2 (a) Tissues: the red square indicates the position of the corresponding patch in the tissue; (b) Patches: the position indicated by the black arrow is the melanoma cells and the nevus cells; (c) Grayscales; (d) Heatmaps: the morphology of melanoma cells is not in regularized shapes, while nevus cells are relatively obvious and regularly arranged; (e) Overlays: multiple layers of data using overlay features to make the ROI more significant.

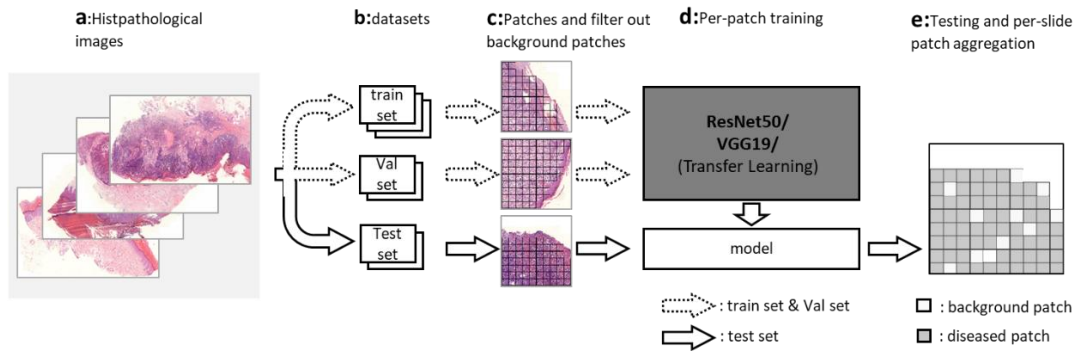

Figure 3 Our classification technique is a kind of patch-based deep CNN.    Data flow is from left to right: (a). slides of diseases (for example, melanoma) was collected in our dataset; (b). slides were then separated into a training (70%), a validation (15%) and a test set (15%); (c). slides were patched by nonoverlapping 256×256 windows, omitting those with over 40% background; (d). the ResNet50 and Vgg19 were used by transferring learning using training set and validating set. (e). classifications were performed on patches from an independent test set, and finally results were aggregated to introduce the AUC statistics.

### 4.2.3. Melanoma from Each Nevus in 40X

In order to test the difficulty of the each performance between melanoma and each kind of nevus, we separately compare three binary classifications, such as melanoma vs. intradermal nevus, melanoma vs. compound nevus, and melanoma vs. junctional nevus. The magnification is set in 40X. Data sizes are 30012 patches of melanoma and 70003 patches of every kind of nevus. Table 3 gave the results of comparison. Again, the values of ResNet50 are better than the value of Vgg19. We also found that relatively the hardest distinction represented by the value of F1 (0.89) lies in the classification between MM and IN of Vgg19, and the easiest one (0.98) lies in MM and JN of ResNet50, which are highlighted with expressed in italics.

Table 3 Performance comparison between melanoma and every kind of nevus in 40X magnifications. The higher value is better. The results highlighted with black bold show the best performance in our dataset.

| | Model | F1 | Sensitivity | Specificity | AUC |
|---|---|---|---|---|---|
| **MM vs. IN** | **ResNet50** | **0.93** | **0.92** | **0.98** | **0.93** |
| | **Vgg19** | *0.89* | 0.86 | 0.97 | 0.88 |
| **MM vs. CN** | **ResNet50** | **0.97** | **0.97** | **0.98** | **0.95** |
| | **Vgg19** | 0.94 | 0.91 | 0.98 | 0.91 |
| **MM vs. JN** | **ResNet50** | ***0.98*** | **0.98** | **0.99** | **0.97** |
| | **Vgg19** | 0.93 | 0.94 | 0.97 | 0.95 |

Table 4 Performance comparison of three kind of nevus in 40X magnifications. The higher value is better. The results highlighted with black bold show the best performance in our dataset.

| | Model | F1 | Sensitivity | Specificity | AUC |
|---|---|---|---|---|---|
| **IN** | ResNet50 | **0.90** | **0.95** | **0.92** | **0.98** |
| | Vgg19 | 0.83 | 0.88 | 0.88 | 0.95 |
| **CN** | ResNet50 | **0.89** | **0.84** | **0.97** | **0.98** |
| | Vgg19 | 0.81 | 0.78 | 0.93 | 0.95 |
| **JN** | ResNet50 | **0.95** | **0.95** | **0.98** | **0.98** |
| | Vgg19 | 0.87 | 0.85 | 0.94 | 0.97 |

### 4.2.4. Classification of Three Nevi

Last comparison is designed to find the answer of classification of three kinds of nevus which are shown in Table 4. The magnification is set in 40X. Data sizes consist of 33336 patches of each kind of nevus. Table 4 gave the results of comparison. ResNet50 got the best value in four metrics.

### 4.3. Interpretation of The Results

CAM is a visualization of the spatial pattern represented in the trained CNN, which generates a heat map of class activation. This generated heat map is a two-dimensional fractional grid associated with a particular output category. Each position of any input image is calculated, which indicates how important each position is to that category. The generated heat map is a two-dimensional fractional grid associated with a particular output category. Each position of any input image is calculated, which indicates how important each position is to that category. For example, for the trained neural network that distinguishes between melanoma and nevi, for each input image, a heat map of melanoma can be generated by CAM visualization, indicating that every part of the image is similar to melanoma's features. Heat maps of nevi can also be generated, indicating how similar every part of the image is to nevus' features.

As shown in Figure 2, the morphological difference between the melanoma and nevus in generated heat maps clearly shows the difference in morphology between melanoma cells and nevus cells. The irregular shape of melanoma cells is present, and the nevus cells are distinctly shaped and regularly distributed. At the same time, the network able to identify key areas of visual recognition through the multiple layers of data using overlay features. These areas almost overlap with the pathologist's region of interest.

### 5. Conclusions

We presented interpretable classification from skin cancer histology slides using deep learning by a retrospective multicancer study. Compared to existing works, large-scale patches of histopathology slides in different magnifications are used to provide morphological diversity. Both classic CNNs architectures are applied to take classifications and achieve much better performance than existed studies. Different cell morphology between melanoma and nevi are illustrated by the visualization of CNN representations and ROI are also located. In the future, we will explore more ways like combining our dataset with different kind of data, such as dermoscopic images, and optimize the training process to simulate the actual clinical diagnosis process by multimodal learning.

Acknowledgment

The authors would like to thank the NCI's Genomic Data Commons for open access to histopathological slides.